# Silicon-chip-based mid-infrared dual-comb spectroscopy

Mengjie Yu[1,2], Yoshitomo Okawachi[1], Austin G. Griffith[3], Nathalie Picqué[4,5,6], Michal Lipson[7] & Alexander L. Gaeta[1]

[1]*Department of Applied Physics and Applied Mathematics, Columbia University, New York, NY 10027, USA*

[2]*School of Electrical and Computer Engineering, Cornell University, Ithaca, NY 14853, USA*

[3]*School of Applied and Engineering Physics, Cornell University, Ithaca, NY 14853, USA*

[4]*Max-Planck-Institut für Quantenoptik, Hans-Kopfermann-Str. 1, 85748 Garching, Germany*

[5]*Ludwig-Maximilians-Universität München, Fakultät für Physik, Schellingstr. 4/III, 80799 München, Germany*

[6]*Institut des Sciences Moléculaires d'Orsay (ISMO), CNRS, Univ. Paris-Sud, Université Paris-Saclay, F-91405 Orsay, France*

[7]*Department of Electrical Engineering, Columbia University, New York, NY 10027, USA*

**Abstract**

The development of a spectroscopy device on a chip that could realize real-time fingerprinting with label-free and high-throughput detection of trace molecules represents one of the "holy grails" of sensing. Dual-comb spectroscopy (DCS) in the mid-infrared is a powerful technique offering high acquisition rates and signal-to-noise ratios through use of only a single detector with no moving parts. Here, we present a nanophotonic silicon-on-insulator platform designed for mid-infrared (mid-IR) DCS. A single continuous-wave low-power pump source generates two mutually coherent mode-locked frequency combs spanning from 2.6 μm to 4.1 μm in two silicon microresonators. A proof-of-principle experiment of vibrational absorption DCS in the liquid phase is achieved acquiring spectra of acetone spanning from 2870 nm to 3170 nm at 127-GHz (4.2-cm$^{-1}$) resolution. These results represent a significant step towards a broadband, mid-IR spectroscopy instrument on a chip for liquid/condensed matter phase studies.

**Introduction**

Dual-comb spectroscopy (DCS)[1-20] is a non-intrusive absorption spectroscopy technique that measures the time-domain interference between two frequency combs of slightly different line spacing. This allows the absorption spectrum to be converted from the optical to the radio-frequency (RF) domain that can be detected rapidly using a single detector. This is particularly critical in the mid-infrared (mid-IR) domain in which sensitive, fast detector arrays remain elusive. The potential of DCS is clear for gas-phase studies in the near-infrared spectral region where DCS shows higher measurement speed, resolution and accuracy than state-of-the-art Michelson-based Fourier transform spectrometers[1]. In recent years, microresonator-based frequency combs have emerged as an attractive compact and broadband source emitting equidistant phase-coherent lines with a large line spacing using a single continuous-wave (CW) pump laser[18,19,21-29]. Microresonator-comb systems could be attractive for spectroscopy in the mid-IR, where sources remain under development[2,4,7,8,10-12,14,16,17]. In the mid-IR region, the absorption strengths of molecular transitions are typically 10 to 1,000 times greater than those in the visible or near-IR, offering the potential to identify the presence of substances with extremely high sensitivity and selectivity. As pointed out by Ideguchi, T. et al.[6], frequency combs of large line spacing (about 100 GHz) and very broad span (approaching an octave or even broader) are required for efficient dual-comb spectroscopy in the liquid or solid-state phases. As such, microresonator-combs represent unique tools for exploring the potential of new approaches to vibrational spectroscopy, in applications where traditional combs based e.g. on modelocked laser systems are not suited. Recently, Suh, et al.[18], demonstrated a dual-comb in the telecommunication region covering 60-nm (6-THz) bandwidth in silica whispering-gallery devices. However, the silica platform cannot be translated into the mid-IR due to inherently high material losses, and the system relies on the use of two separate pump lasers.

In this work, we present a CMOS-compatible silicon-based, chip-scale mid-IR dual-comb spectrometer that meets the requirements for vibrational DCS and mid-IR molecular fingerprinting in the condensed phase. Two mutually coherent modelocked frequency combs are generated using a single CW laser spanning from 2.6 μm to 4.1 μm. Thermal control and free-carrier injection allow for independent modelocking of each comb and for tuning of the dual-comb parameters. The large line spacing of the combs (127 GHz) and its precise tuning over tens of MHz, which are unique features of chip-scale comb generators, are exploited for a proof-of-principle experiment of vibrational absorption DCS of acetone. This work represents a critical advance for chip-based linear DCS for liquid/condensed matter phase studies, which would find a wide range of applications in chemistry, bio-medicine, material science, and industrial process control.

With further development, it holds promise for real-time and time-resolved spectral acquisition on nanosecond time scales.

**Results**

**Generation of the dual-comb source**

The experimental setup is shown in Fig. 1. We use two silicon microresonators that have 100-μm radii and are dispersion engineered to have anomalous group-velocity dispersion beyond 3 μm for the fundamental TE mode, similar to Griffith, *et al.*[27]. A CW optical parametric oscillator (100-kHz linewidth) emitting at 3 μm simultaneously pumps two microresonators with slightly different line spacings. The two generated combs are combined at a beamsplitter and sent to a photodetector (bandwidth of 250 MHz) connected to an RF spectrum analyzer. Integrated PIN diodes, located around the resonators, are operated at a reverse-bias voltage of -15 V to sweep out the free carriers (FC) generated from three-photon absorption (3PA)[25]. We generate a modelocked mid-IR frequency comb in both microresonators simultaneously by tuning the pump laser into the cavity resonances. A thermoelectric cooler (TEC) is used to control the temperature of each silicon device independently in order to compensate the initial frequency difference between the two pump resonances, and for coarse tuning of the difference in repetition frequencies between the two combs Δ$f_{rep}$ to lie within our detector bandwidth. Future implementations could allow for fully integrated microheaters[29], which would achieve even more precise control of the line spacing of each of the microresonators while drawing little power. The mutual coherence between the two combs is established by sharing the same pump laser and from the inherent modelocking mechanism of microresonator-based combs. The RF beatnotes of the dual-comb output appear at frequencies $f_N = N*\Delta f_{rep}$, where $N$ is an integer. Figure 1 (inset) shows the mapping of the optical spectrum to the RF dual-comb spectrum where the shorter and longer wavelength sides of the pump are mapped to the same RF domain, which means appropriate long-pass/short-pass filters are needed to access either side of the optical spectrum relative to the pump frequency. Shifting the pump frequency of one of the microresonators, e.g., with an acousto-optic modulator, would avoid such aliasing, as already demonstrated with electro-optic-modulator-based dual-comb spectroscopy[9,13,15,16].

Figure 2a shows the generated spectrum of one of the combs measured by a Michelson-based Fourier transform infrared spectrometer (M-FT). The spectrum consists of 305 comb lines with a spacing $f_{rep}$ = 127 GHz and spans 2.6 – 4.1 μm, which is the region of the fundamental CH, NH and OH stretching modes in molecules. The pump powers for each microresonator are 80 and 50 mW, and the pump-to-comb conversion efficiencies are each >30 %. The power of each comb line varies from 2.5 μW to 2 mW in the range of 2.8 – 3.2 μm and the power variation is due to the modulation in the optical spectrum from multiple solitons generated within one cavity roundtrip[28,29]. Since the cavity linewidth ($10^5$ $Q$-factor) is broader than the detector bandwidth, the sharp comb linewidth when in the modelocked state is crucial for resolving the RF beatnotes of the two combs. The transition to the modelocked state is determined by the observation of a step in the optical transmission[24], an abrupt increase in the 3PA-induced FC current[28], and the transition to a low RF noise state as shown in Fig. 2b[22].

While the M-FT spectra of the two individual and combined modelocked frequency combs are shown in Fig. 3a, the RF dual-comb spectrum is plotted in Fig. 3b where the intensity profile agrees well with the product of the amplitudes of the electric fields of two optical frequency combs within the detection range. The achieved difference between the line spacings $\Delta f_{rep}$ is 12.8 MHz, corresponding to a frequency compression factor from the optical to RF domain of $f_{rep}/\Delta f_{rep}$ (~10,000). The minimum time required to resolve the RF comb lines is $1/\Delta f_{rep}$ (~78 ns), indicating the potential for a rapid single-shot measurement. The measured linewidth of the 25th RF comb line [Fig. 3b inset] is < 100 kHz at a RF resolution bandwidth of 40 kHz, which corresponds to a mutual coherence time between the two combs of >10 μs. The frequency jittering of the optical frequency combs is dominated by the pump laser[28]. Therefore, the coherence between the two generated combs is drastically improved because the effect of the pump noise is expected to be significantly minimized by sharing the same pump with the two microresonators.

**Tunability of the dual-comb source**

While the spectral window of 2.6 – 4.1 μm (73 - 115 THz) can be mapped into an RF window of 2.5 GHz with $\Delta f_{rep}$ = 12.8 MHz , our measurement spectral range is currently 3 – 3.12 μm, which is largely limited by our detector bandwidth of 250 MHz. Our detection range can be extended to cover the entire spectral window by using a faster detector or by controlling the repetition rate of the two combs to achieve a smaller $\Delta f_{rep}$ (e.g., 1 MHz). Coarse tuning of the $\Delta f_{rep}$ can be achieved by thermally tuning the resonances of one microresonator by one or more FSR's. In our case, we found that the $\Delta f_{rep}$ is changed by ~126 MHz by moving to the next resonance in one of the microresonators, which is about 1/1000 of the FSR. Additionally, we achieve fine tuning of $\Delta f_{rep}$. Figure 4 shows the repetition rate tuning of our dual-comb system. We observe that $\Delta f_{rep}$ can be finely tuned by >10 MHz by simply changing the pump-cavity detuning while maintaining modelocking in both microresonators. Moreover, $\Delta f_{rep}$ is dependent on both the FC dispersion effect and on the thermo-optic effect. By individually controlling the reverse-bias-voltage applied on the PIN junctions, the line spacings of the microresonators can be finely controlled independently such that smaller $\Delta f_{rep}$ (~ 1 MHz) can be achieved in the near future. This tuning technique of the RF line spacing ($\Delta f_{rep}$) provides flexibility in achieving an optimal refresh rate of the measurement over a desirable spectral range and in further compressing the needed RF window.

**Time-domain interferogram and dual-comb spectroscopy**

For our spectroscopy experiments, the spectra are obtained by taking the Fourier transform of the time-domain interference signal, which is measured using a photodetector and a fast real-time oscilloscope with a sampling rate of 80 Gbps. The difference in the line spacing between the two 127-GHz combs is tuned to be 39 MHz. A time-domain interferogram with a measurement time of 2 μs is shown in Fig. 5a which displays a periodic waveform that repeats every 25.6 ns, which corresponds to the inverse of the difference in comb line spacing. The waveform shows good reproducibility and has multiple peaks within one period due to the interference of multiple solitons generated in both microresonators within a cavity roundtrip. The Fourier-transform of the interferogram reveals a RF spectrum (Fig. 5b) where 30 comb lines are resolved with a line spacing of 39 MHz. The observed comb linewidth is 0.5 MHz, limited by the recording time of 2 μs. We define the SNR based on the intensity of the comb line divided by the standard deviation of the noise baseline between two comb lines. The average of the signal-to-noise ratio (SNR) exceeds 6000, and the variation of the SNR over the comb teeth is due to multiple soliton formation. The average SNR per unit of time exceeds $4 \times 10^6$ $s^{-1/2}$, and its product with the number of spectral elements exceeds $10^8$ $s^{-1/2}$, which is higher than the previously reported value of $10^6$ $s^{-1/2}$ for a mid-IR dual comb measurement[4].

We illustrate the potential of our dual-comb spectrometer for broadband vibrational spectroscopy of liquid samples with a proof-of-principle absorption measurement. We insert a 100-μm thick cuvette filled with neat acetone in the combined arm of the dual-comb interferometer. Here, we utilize two different bandpass filters to access each side of the optical spectrum with respect to the pump wavelength. The absorption spectra for each side of the pump are measured with an acquisition time of 2 μs using a faster photodetector of 1-GHz bandwidth, corresponding to a spectral window of 2870 – 3170 nm. The dual-comb absorbance and transmittance (inset) spectra are shown in Fig. 6 at a spectral resolution of 127-GHz (4.2-cm$^{-1}$) resolution. The transmittance is calculated as the ratio between the spectrum with and without the cuvette which are measured sequentially. The absorbance is the logarithm of the transmittance. We compare the results to the absorption spectrum we measured using a Fourier transform spectrometer (Bruker Vertex 70) equipped with a globar, a $CaF_2$ beam-splitter and a InSb detector. The instrumental resolution is 0.5 $cm^{-1}$ (16 GHz), which is much narrower than the spectral bands of the liquid molecular sample. The two spectra are in reasonable agreement. However, we observe a low SNR of 25, as determined on the higher transmittance side between 3000 nm to 3100 nm. The large noise comes from the fact that the two spectra with and without the cuvette are measured at different times, resulting in deviations primarily due to the long-term stability of the interference signal. The free-space interferometer is sensitive to any mechanical vibrations and environmental change which adds baseline noise to a single spectrum at relatively low frequencies. This long-term stability can be improved by comparing the two spectra with and without the sample at the same time. Additional deviations may arise from the cross-talk of the two bandpass filters and averaging over multiple acquisitions can reduce the root-mean-square deviation but at a cost of a longer acquisition time.

**Discussion**

Our dual-comb spectrometer using silicon-microresonator-based frequency combs will readily achieve spans from 2600 to 4100 nm by reduction of the difference in repetition frequencies. A difference $\Delta f_{rep}$= 3.3 MHz would map the 305 comb lines within the 1-GHz bandwidth of our fastest detector. Spectra of 42-THz bandwidth at 127-GHz resolution will then be acquired at a refresh time of 0.3 μs. For comparison, the fastest M-FT interferometer[31] measures interferograms at a resolution of 270 GHz (9 $cm^{-1}$) with a refresh rate of 80 kHz (refresh time: 13 μs). Moreover, our refresh rate is fundamentally limited by the Nyquist limit to 200 MHz. As losses in the silicon nanowaveguides continue to decrease, we expect that microresonators with smaller FSR's will be achievable which will allow for higher spectral resolution. Real-time averaging with field-programmable gate arrays may boost the sensitivity to weak absorptions of trace molecules. With future developments to our CMOS-compatible platform, such as implementation of quantum cascade lasers as pump sources, broader spans and access to other ranges of the molecular fingerprint region can also be envisioned. With continued progress to instrumentation such as mid-IR detectors and digitizers, we believe this system will evolve into a spectroscopy laboratory on a chip for real-time vibrational sensing on the nanosecond time scale.

**Data Availability**

The data that support the plots within this paper and other findings of this study are available from the corresponding author upon reasonable request.

**Acknowledgments**

We thank Jun Ye from JILA, University of Colorado for useful discussions. We acknowledge support from Defense Advanced Research Projects Agency (W31P4Q-15-1-0015), the Air-Force Office of Scientific Research (FA9550-15-1-0303), and National Science Foundation (ECS-0335765, ECCS-1306035). This work was performed in part at the Cornell Nano-Scale Facility, a member of the National Nanotechnology Infrastructure Network, which is supported by the National Science Foundation (NSF) (grant ECS-0335765).

**Author contributions**

M.Y. and Y.O. prepared the manuscript in discussion with all authors. M.Y. and Y.O. designed and performed the experiments. A.G.G. fabricated the devices. N.P. contributed to preliminary data analysis. M.L. and A.L.G. supervised the project. All authors discussed the results and commented on the manuscript.

**Additional information**

The authors declare no competing financial interest.

**Corresponding author**

Correspondence: Author to whom all correspondence should be directed.

Alexander L. Gaeta – alg2207@columbia.edu

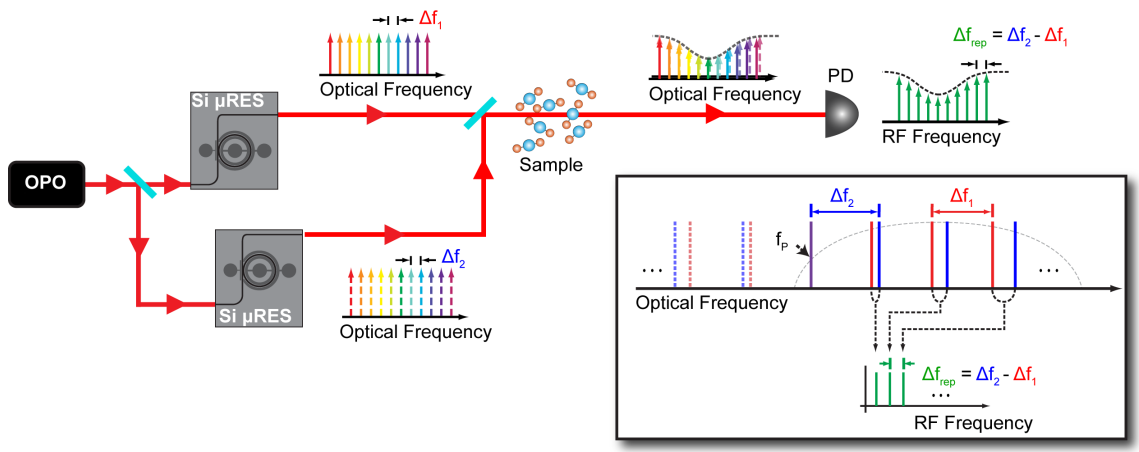

**Figure 1 | Schematic for dual-comb absorption spectroscopy.** Experimental setup for our dual-comb source. A continuous-wave optical parametric oscillator pumps two separate silicon microresonators, which generate two modelocked combs. The output is combined and sent to a photodiode for RF characterization. Inset: Schematic for single-pump operation and mapping from optical to RF domain. $\Delta f_1$ and $\Delta f_2$ are the repetition frequencies of two optical frequency combs. $\Delta f_{rep} = \Delta f_2 - \Delta f_1$ is the difference in repetition frequencies. PD, photodiode; Si μRES, silicon microresonator; OPO, optical parametric oscillator.

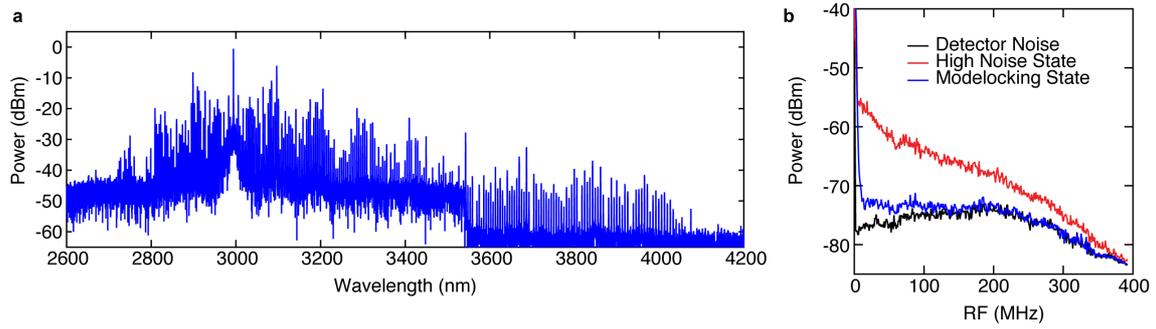

**Figure 2 | Silicon microresonator-based dual-comb source.** (a) A spectrum of one of the generated combs measured using a Michelson-based Fourier transform infrared spectrometer (M-FT). The spectral range is from 2.6 µm to 4.1 µm. The resolution is 7 GHz (0.25 cm$^{-1}$). (b) RF-noise characterization of the generated comb. The plot shows the reduction in RF amplitude noise corresponding to modelocking.

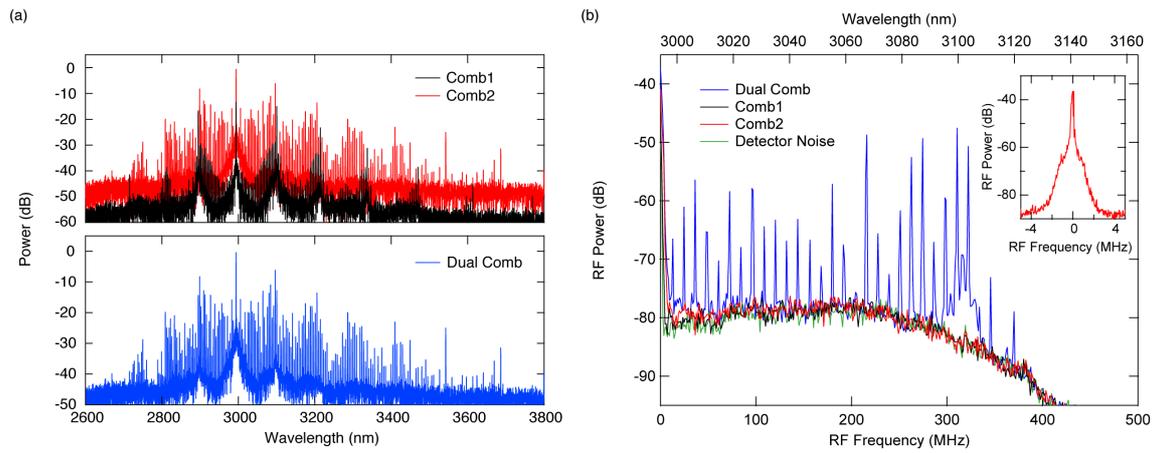

**Figure 3 | Characterization of dual-comb source.** (a) Top: M-FT spectra for each modelocked comb. Bottom: Combined M-FT spectrum. (b) RF spectrum from the dual-comb interferometer. Plot shows RF spectra for dual-comb (blue), each separate modelocked comb (black and red), and detector noise background (green). Inset: Characterization of the 25th RF beatnote in (b).

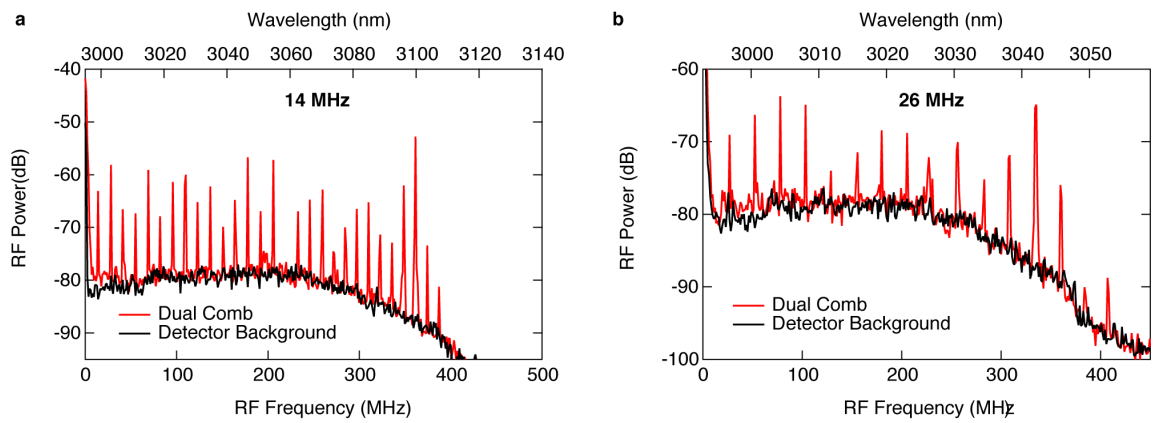

**Figure 4 | Repetition rate tuning of the dual-comb source.** The frequency spacing of the dual-comb source is dictated by the spacing of each of the modelocked combs. The plot shows a 14 MHz spacing (left) and a 26 MHz spacing. The spacing is tuned by adjusting the TECs to change the resonance position of the microresonators.

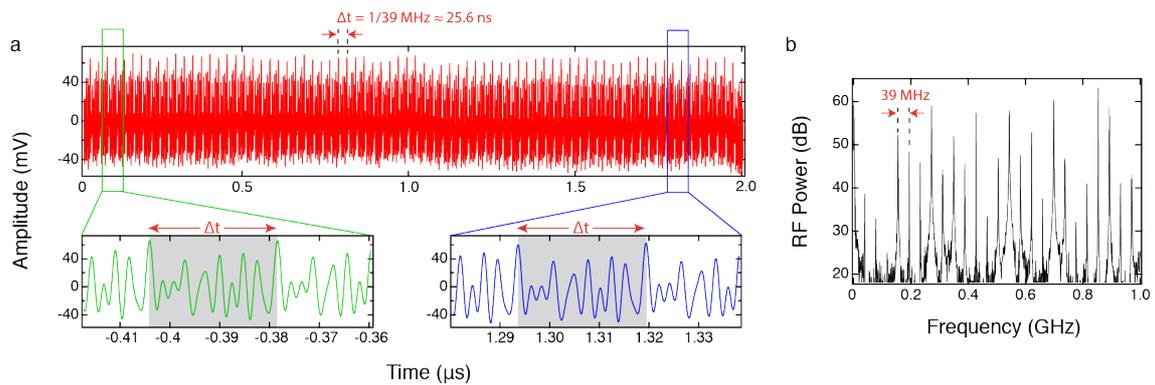

**Figure 5 | Experimental interferogram and spectrum.** (a) Time-domain interferogram over a measurement time of 2 µs. The waveform repeats with a period of 25.6 ns, which is the inverse of the difference in comb line spacing (39 MHz). The waveform of one period is shaded and demonstrates good reproducibility. The multiple bursts within one period indicate that multiple solitons are generated in each microresonator within one cavity roundtrip. (b) Fourier-transformed spectrum of the time-domain interferogram in (a), on a logarithmic scale with 25 resolved lines and an average signal-to-noise ratio exceeding 6000. Modulation in the spectrum is also due to both comb operating in multiple soliton regime.

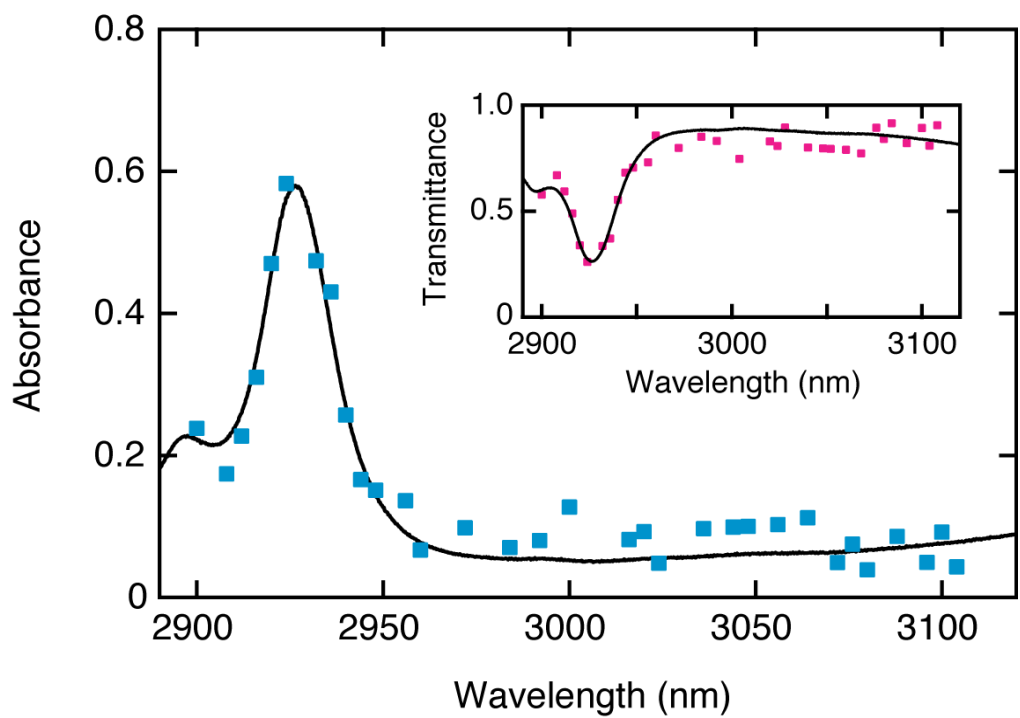

**Figure 6 | Absorption spectroscopy over a short measurement time.** Two different bandpass filters are used to access the two sides of the dual-comb spectrum that are symmetric with respect to the pump wavelength of 3 µm. The measurement time is 2 µs for each side of the dual-comb spectrum. The transmittance (inset) is calculated as the ratio between the spectrum with the cuvette and the spectrum without the cuvette measured sequentially. The absorbance is the logarithm of the transmittance. The results are compared to the absorption measurement using a M-FT spectrometer equipped with a globar. A SNR of 25 is extracted between 3000 nm to 3100 nm largely due to a long-term stability issue of the interferogram and could be improved by comparing the two spectra with and without the sample at the same time.